# A Pre-trained Foundation Model Framework for Multiplanar MRI Classification of Extramural Vascular Invasion and Mesorectal Fascia Invasion in Rectal Cancer

Yumeng Zhang[1], Shruti Atul Mali[1], Danial Khan[1], Sina Amirrajab[1], Eduardo Ibor-Crespo[3], Ana Jimenez-Pastor[3], Gloria Ribas[4], Silvia Flor-Arnal[4], Marta Zerunian[6], Christophe Aubé[7,8], Luis Marti-Bonmati[4,5], Zohaib Salahuddin[1*], Philippe Lambin[1,2*]

[1] The D-Lab, Department of Precision Medicine, GROW - Research Institute for Oncology and Reproduction, Maastricht University, 6220 MD Maastricht, The Netherlands

[2] Department of Radiology and Nuclear Medicine, GROW - Research Institute for Oncology and Reproduction, Maastricht University, Medical Center+, 6229 HX Maastricht, The Netherlands

[3] Research & Frontiers in AI Department, Quantitative Imaging Biomarkers in Medicine, Quibim SL, Valencia, Spain

[4] Biomedical Imaging Research Group, La Fe Health Research Institute, Valencia, Spain

[5] Medical Imaging Department, La Fe University and Polytechnic Hospital, Valencia, Spain

[6] Radiology Unit, Department of Surgical and Medical Sciences and Translational Medicine, Sapienza University of Rome, Sant'Andrea Hospital, 00189 Rome, Italy

[7] Laboratoire HIFIH, Université d'Angers, SFR ICAT 4208, Angers 49000, France

[8] Department of Radiology, CHU Angers, Angers 49000, France

*Equal contribution

## Abstract

**Objectives** Accurate MRI-based identification of extramural vascular invasion (EVI) and mesorectal fascia invasion (MFI) is crucial for risk-stratified rectal cancer treatment. However, subjective visual assessment and inter-institutional variability limit diagnostic consistency. This study developed and externally evaluated a multi-centre, foundation model-driven framework that automatically classifies EVI and MFI on axial and sagittal MRI.

**Methods** A total of 331 pre-treatment rectal cancer T2-weighted MRI scans from three European hospitals were retrospectively recruited. A self-supervised frequency domain harmonization strategy was applied to reduce scanner variability. Three classifiers, SeResNet, the universal biomedical pretrained model (UMedPT) with a multilayer perceptron head, and a logistic-regression variant using frozen UMedPT features (UMedPT_LR), were trained (n=265) and tested (n=66). Gradient-weighted class activation mapping (Grad-CAM) visualized model predictions.

**Results** UMedPT_LR achieved the best EVI performance with multiplanar fusion (AUC=0.82, test set). For MFI, UMedPT trained on axial harmonized images yielded the highest performance (AUC=0.77). Both tasks outperformed the CHAIMELEON 2024 benchmark (EVI: 0.82 vs 0.74; MFI: 0.77 vs 0.75). Harmonization enhanced MFI classification, and multiplanar fusion further boosted EVI performance. Grad-CAM confirmed biologically plausible attention on peritumoral regions (EVI) and mesorectal fascia margins (MFI).

**Conclusion** The proposed foundation model-driven framework, leveraging frequency domain harmonization and multiplanar fusion achieves state-of-the-art performance for automated EVI and MFI classification on MRI, demonstrating strong generalizability across multiple centers.

## Introduction

Rectal cancer accounts for approximately one-third of colorectal cancer cases and remains a major cause of cancer-related mortality worldwide[1–3]. Accurate staging is essential for tailoring treatment strategies and improving patient outcomes[4]. Among key prognostic markers, extramural vascular invasion (EVI) and mesorectal fascia invasion (MFI) play critical roles in guiding therapeutic decisions. EVI refers to the extension of tumor cells beyond the muscularis propria into perirectal vessels and is strongly associated with local recurrence, distant metastasis, and poor overall survival. Its presence can influence management even in node-negative patients, supporting the recommendation of neoadjuvant chemoradiotherapy (nCRT) to reduce recurrence risk[5]. MFI describes tumor involvement of the mesorectal fascia, the connective tissue envelope surrounding the rectum. Identification of MFI indicates a high risk of an involved circumferential resection margin (CRM), prompting guidelines to recommend nCRT to downstage the tumor and enhance the likelihood of a clear CRM during surgery[6].

MRI provides direct visualization of EVI and MFI and is considered the gold standard for locoregional staging of rectal cancer[7]. On T2-weighted MRI, EVI appears as irregular extensions beyond the muscularis propria into perirectal vessels.[8] While axial images best depict these vascular protrusions[9], sagittal images provide complementary information on the longitudinal extent of vascular involvement[10]. By contrast, MFI is characterized by loss of the normally sharp mesorectal fascia, reflecting tumor extension toward or into this boundary[11]. Although sagittal images are less precise for defining fascial margins, they remain valuable for assessing craniocaudal tumor spread and its relationship to the anal canal and pelvic floor[12]. Collectively, these perspectives underscore the relevance of multiplanar MRI assessment for accurate evaluation of both EVI and MFI.

While experienced radiologists are able to identify these features on T2-weighted MRI, interpretation is subjective, varies between institutions, and becomes particularly variable in borderline cases. These limitations highlight the need for artificial intelligence (AI)-based approaches that provide objective and reproducible assessments, thereby supporting consistent staging decisions[13].

Radiomics and deep learning have emerged as promising strategies for extracting high-dimensional imaging features imperceptible to the human eye[14–16]. These methods have demonstrated clinical utility in predicting treatment response, nodal metastasis, and survival in rectal cancer patients[17]. However, standard deep learning models are highly sensitive to inter-institutional variability in scanner type and acquisition protocols[18, 19], particularly in multi-center studies[20]. To address this challenge, harmonization methods such as ComBat[21] or CycleGAN[22] have been proposed, whereas more recent frequency domain and self-supervised approaches aim to learn modality-invariant representations with minimal supervision[23].

Recently, foundation models trained on large and heterogeneous medical image datasets have emerged as a promising approach to mitigate these limitations by enabling generalizable and transferable feature learning[24]. For instance, RETFound[25] for retinal imaging and BEPH[26] for histopathology demonstrate how large-scale pretraining can yield transferable features. However, these models remain limited to modality-specific applications. The Universal Biomedical Pretrained Model (UMedPT[27]) is a cross-domain foundation model designed to support diagnostic tasks across diverse biomedical imaging modalities. Its potential has not yet been explored for rectal MRI classification tasks.

In this study, we propose a foundation model–based framework for automated classification of EVI and MFI in rectal cancer using axial and sagittal T2-weighted MRI from multiple vendors and centers. Our approach leverages the UMedPT foundation model alongside a logistic regression classifier applied to deep features extracted from UMedPT (UMedPT_LR). To mitigate inter-institutional variability, we incorporate a self-supervised frequency domain harmonization strategy, and to enhance interpretability, we apply Gradient-weighted class activation mapping (Grad-CAM) to visualize the spatial rationale of model predictions.

## Methods

### Workflow overview

An overview of the study workflow is presented in **Fig.1.**.

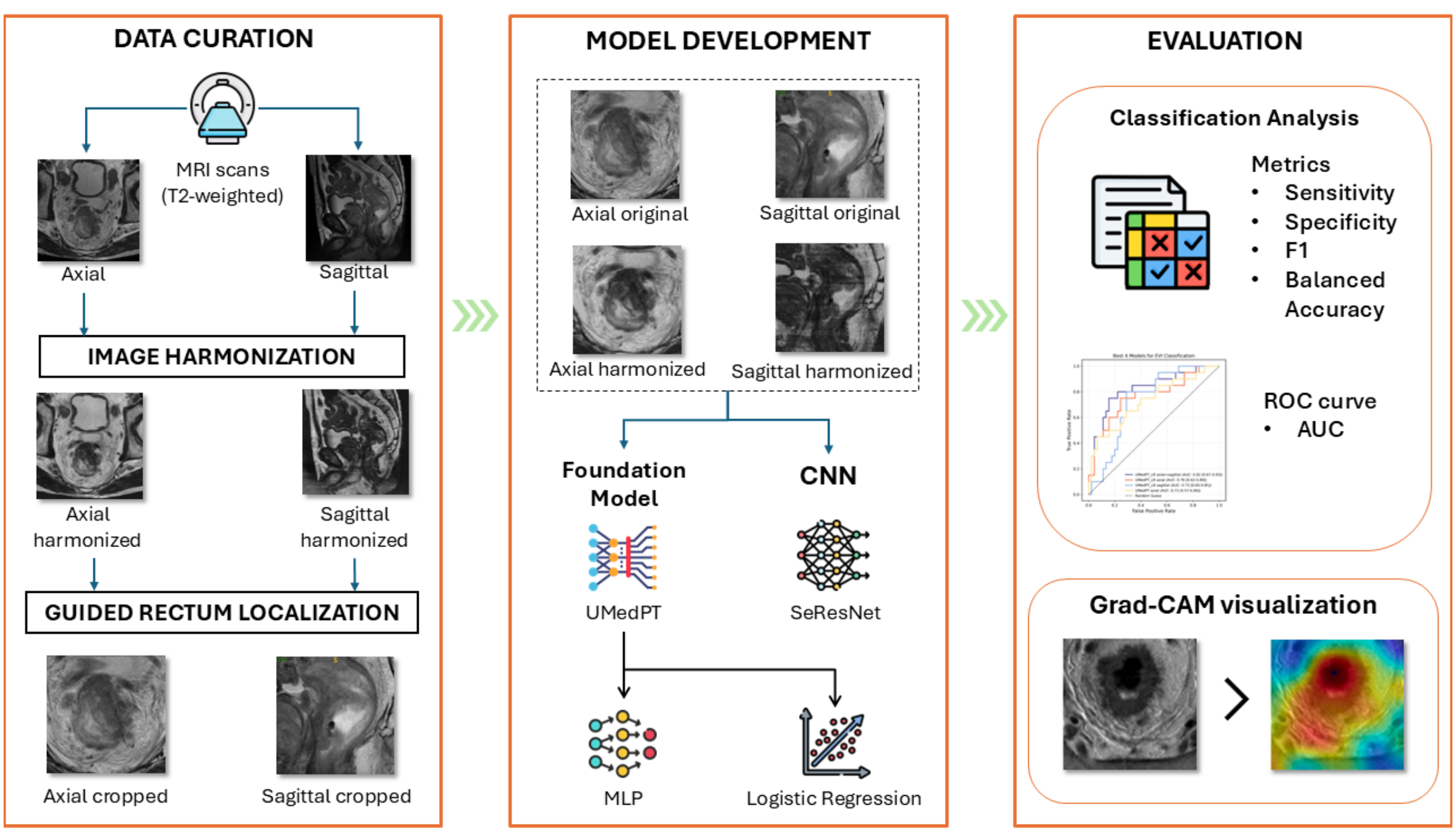

Fig.1. Workflow of this study. The process comprises three main stages: **(i)** data curation, including acquisition of axial and sagittal T2-weighted MRI images, image harmonization to reduce inter-scanner variability, and guided localization of the rectum; **(ii)** model development, where both the original and harmonized images were processed using a foundation model (UMedPT) and a convolutional neural network (SeResNet), followed by classification using either a multilayer perceptron

(MLP) or logistic regression; and **(iii)** evaluation, involving classification performance analysis with multiple metrics (sensitivity, specificity, F1 score, and balanced accuracy), receiver operating characteristic (ROC) curves with area under the curve (AUC) computation, and model interpretability assessment via Grad-CAM visualization.

## Data curation

### Dataset

This retrospective study utilized data from the European CHAIMELEON[28] project, a large-scale initiative aimed at developing AI algorithms for cancer diagnosis across multiple tumor types, including lung, colon, breast, prostate, and rectum cancers. All imaging and clinical data were processed within the CHAIMELEON platform, which ensured secure handling and compliance with privacy regulations. In this study, we included 331 rectal cancer patients from three hospitals – La Fe University and Polytechnic Hospital (La Fe, n=194), Sant'Andrea University Hospital, Sapienza University of Rome (ULS, n=113), and Centre Hospitalier Universitaire d'Angers (CHU Angers, n=24) – utilizing MRI scanners from mainly two manufacturers – GE (n=221) and Siemens (n=106).

Patient inclusion criteria were: (1) biopsy-confirmed rectal adenocarcinoma; (2) age ≥ 18 years (diagnosis year ≥ 2014); (3) No distant metastasis at baseline (M0); and (4) availability of pre-treatment pelvic MR exam with axial and sagittal T2-weighted images. (5) Patients were followed ≥ 12 months from the first treatment, unless death, recurrence, or progression occurred earlier. Exclusion criteria were absence of baseline pelvic MR (n=12) or incomplete sequences (n=4).

As illustrated in **Fig.2.**, the dataset was split into a train_val set (n=265, 80%) and a test set (n=66, 20%). Within the train_val cohort, 80% (n=212) was used for training and the remaining 20% (n=53) for validation. Ground truth for EVI and CRM involvement, which represent the pathological correlate of MFI, was obtained from histopathology. Two radiologists independently verified tumor localization and excluded non-tumorous structures.

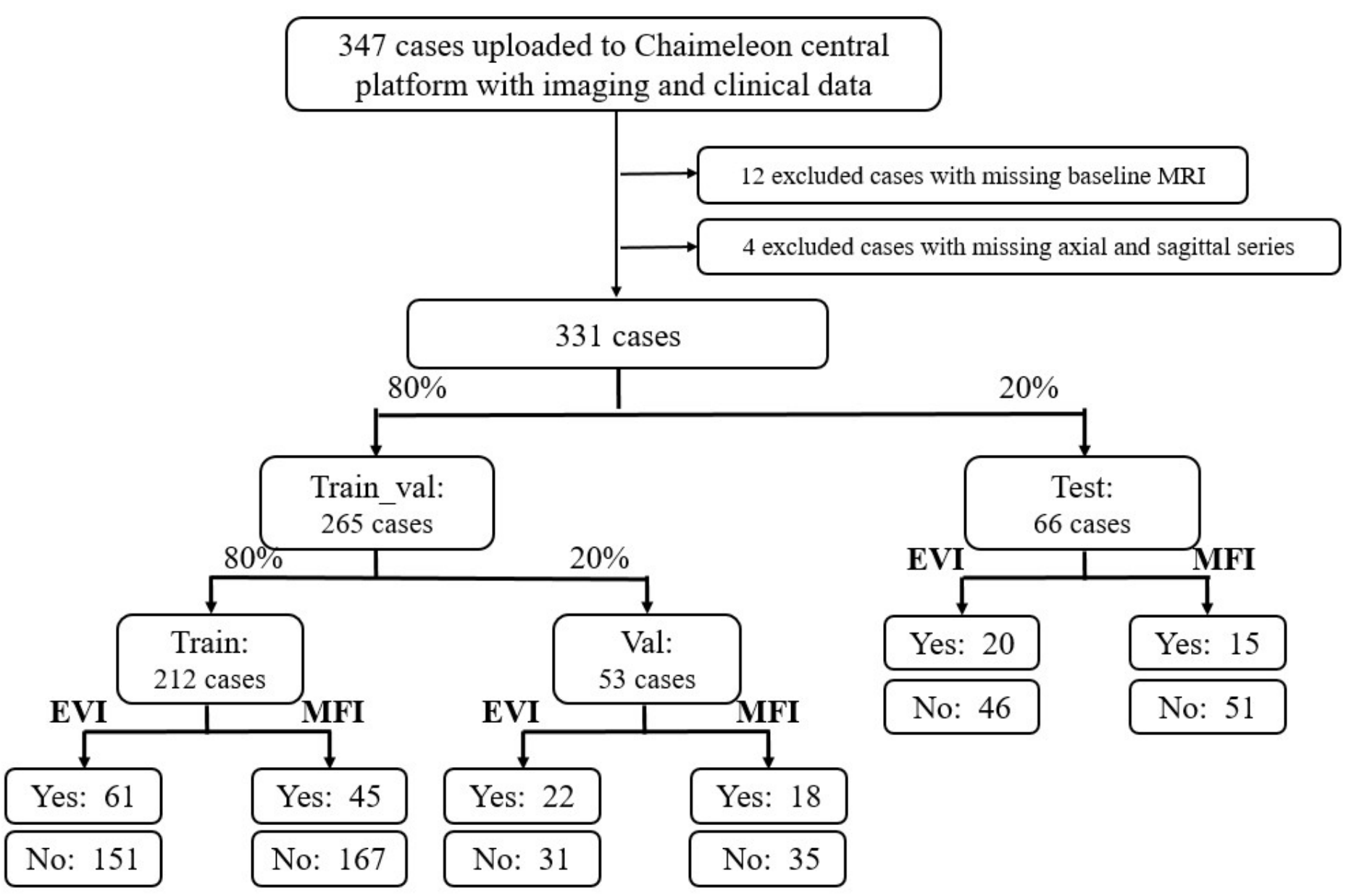

Fig.2. Flowchart of data inclusion, exclusion, and dataset stratification for model development and evaluation

## Image preprocessing and Rectum localization

Data were converted from DICOM to NIfTI and processed with a structured automatic pipeline. Pelvic anatomy was segmented using TotalSegmentator[29]. The colon mask served as the primary cue for rectal localization, with adjacent structures (e.g., bladder) used as auxiliary references. We computed the rectal centroid within the colon mask and applied a fixed-size center-crop patch to both axial and sagittal planes, ensuring consistent rectal coverage and alignment across patients. All series were resampled to a uniform voxel spacing of 0.39 × 0.39 × 3.3 mm to standardize resolution across scans. To reduce inter-patient intensity variation, voxel values were clipped to the 2.5–97.5th percentiles and subsequently Z-score normalized.

## Frequency domain harmonization

To address inter-institutional variability in multi-centre T2w MR dataset, we employed a frequency domain harmonization pipeline developed by QUIBIM (Valencia, Spain) as a part of the CHAIMELEON project. A curated reference set defined the target contrast distribution, synthetic variants were generated by Fourier-domain perturbations, and a self-supervised UNet-based autoencoder learned to reconstruct standardized images from perturbed inputs.

The resulting outputs were used alongside originals in downstream classification. Complete specifications and training schedule are present in **Appendix A1**.

## Model development

### SeResNet

We employed a modified version of Squeeze and Excitation Residual Network (SeResNet)-34 with an anisotropic stem (3×3×1) and early (2,2,1) strides to respect through-plane resolution in pelvic MRI. The network ends with global average pooling and a sigmoid output for binary EVI/MFI classification. Full layer configuration and training settings are in **Appendix A2**.

### UMedPT

UMedPT[27] is a foundation model designed to learn transferable representations across diverse biomedical imaging modalities and tasks. In this study, we adapt UMedPT for the classification of EVI and MFI from T2-weighted MRI pelvic images. As illustrated in **Fig. 3**, the input 3D volume was first decomposed into a stack of 2D slices. Each slice is processed independently through a shared encoder-squeezer module pretrained within the UMedPT framework. The encoder, based on the Swin Transformer architecture, extracts high-level modality-agnostic features and produces a hierarchical representation with 288 tokens, each compressed into a 512-dimensional feature vector by the squeezer. These features are subsequently aggregated by the grouper into a single 512-dimensional representation for classification. This is followed by a Multilayer perceptron (MLP) classifier, consisting of two fully connected layers with ReLU activations and dropout, which outputs binary predictions for EVI or MFI.

During fine-tuning, only the grouper and classifier layers were optimized, while the encoder and squeezer components were frozen to preserve pretrained feature representations. Input images were duplicated across three channels to conform with the encoder's requirements. This architecture enables efficient transfer learning and robust volume-level classification, while minimizing the number of trainable parameters required for domain adaptation.

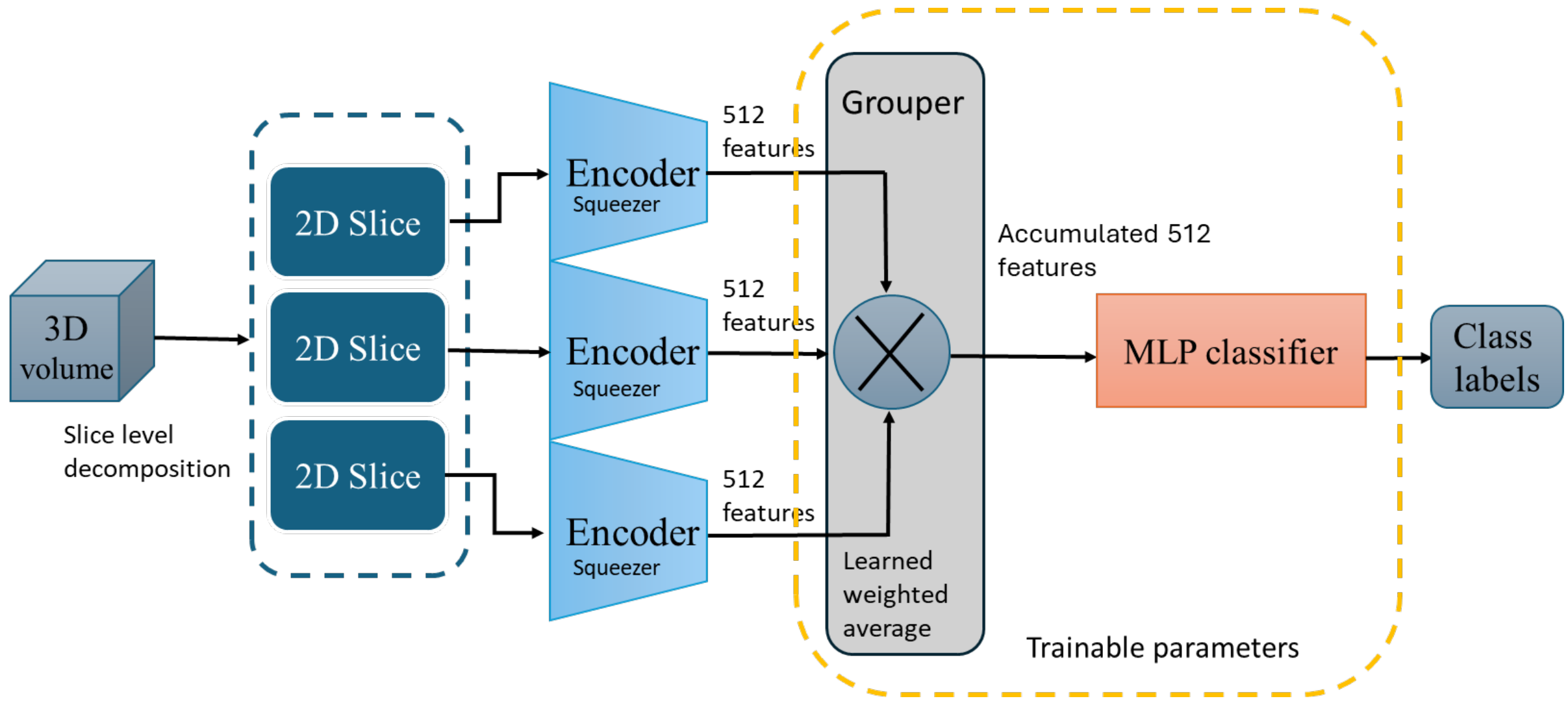

Fig.3. UMedPT architecture for rectal cancer stage classification using multi-slice MRI. Each 3D MRI volume is decomposed into 2D slices, which are individually encoded and compressed via shared encoders and squeezer modules pretrained in UMedPT. The extracted 512-dimensional features from all slices are aggregated using a learned weighting strategy in the Grouper module. The accumulated representation is then passed to a MLP classifier to predict the presence or absence of extramural vascular invasion (EVI) and mesorectal fascia invasion (MFI). During fine-tuning, only the grouper and classifier parameters are updated, while the backbone encoder remains frozen.

### UMedPT_LR and Multiplanar fusion

UMedPT_LR replaces the MLP head with a logistic regression classifier applied to the 512-dimensional UMedPT grouper features. Feature dimensionality was reduced using principal component analysis (PCA) before fitting the logistic-regression model. This design lowers computational cost and enhances interpretability through component-level coefficients.

In the multiplanar fusion setting, we concatenate the 512-dimensional UMedPT grouper features from the axial and sagittal models into a 1024-dimensional vector. This fused representation then undergoes the same pipeline as UMedPT_LR.

### Training strategy

Data augmentation (MONAI[30]) comprised random affine transforms, intensity adjustments, Gaussian noise and smoothing, and MRI-specific artifacts. Detailed settings are provided in **Appendix A3**. All images were rescaled to the 2.5th–97.5th intensity percentile range. A patch size of 192 × 192 × 36 was employed to capture the full lesion context while minimizing interference from surrounding anatomy.

SeResNet was optimized using α-balanced focal loss and trained for 200 epochs using SGD with a learning rate of $1\times10^{-3}$. A warm-up of 10 epochs and ReduceLROnPlateau scheduling were used. UMedPT used a weighted binary cross-entropy loss and AdamW optimizer with a learning rate of $5\times10^{-4}$. Cosine annealing was used for learning rate decay. Both models were trained with a batch size of 8 using half-precision floating-point arithmetic.

## Evaluation

### Evaluation metrics

Model performance was evaluated using AUC as the primary metric, alongside balanced accuracy, sensitivity, specificity, and F1 score to account for class imbalance. Bootstrap resampling (1000 iterations) was used to compute 95% confidence intervals, ensuring robust and reliable estimates. Sensitivity and specificity were computed from binary predictions, obtained by thresholding the predicted probabilities at 0.5.

Sensitivity and specificity are defined as:

$$Sensitivity = \frac{TP}{TP+FN} \tag{1}$$

$$Specificity = \frac{TN}{TN + FP} \tag{2}$$

Balanced accuracy accounts for both sensitivity and specificity, offering a fairer evaluation when class distributions are skewed:

$$Balanced\ Accuracy = \frac{(Sensitivity + Specificity)}{2} \tag{3}$$

The F1 score provides the harmonic mean of precision and recall:

$$F1 = \frac{TP}{TP + \frac{1}{2}(FP + FN)} \tag{4}$$

### Model interpretability

To interpret the spatial rationale underlying model decisions, we applied Gradient-weighted class activation mapping (Grad-CAM) [31] to the trained UMedPT-based classifier.

Specifically, for each axial original T2-weighted MRI scans, the input image was normalized and converted to RGB format. Grad-CAM heatmaps were generated by targeting the norm1 layer of the first SwinTransformerBlockV2 (the hierarchical transformer module within the UMedPT encoder) in the 7th sequential block, which corresponds to high-level contextual features. The resulting heatmaps were normalized and overlaid on the original MRI slices.

### Reporting quality assessment

We evaluated methodological quality using the Radiomics Quality Score 2.0 (RQS 2.0)[32]. This framework provides standardized criteria to assess transparency, reproducibility, and clinical applicability in radiomics and AI-based imaging studies.

# Results

## Patient characteristics

The characteristics of the patient cohort are summarized in Table 1. The median age of the patients was 65 years (range: 28–89 years), and 60.7% (n=201) were male. Extramural vascular invasion (EVI) was positive in 31.1% (n=103) of cases, while mesorectal fascia invasion (MFI) was positive in 23.6% (n=78). Patients were recruited primarily from three hospitals: LaFe (58.6%, n=194), ULS (34.1%, n=113), and CHU Angers (7.3%, n=24). MRI scans were predominantly obtained using GE scanners (66.8%, n=221), followed by Siemens scanners (32.0%, n=106), with a small minority from other manufacturers (1.2%, n=4). The demographic and clinical characteristics were consistent across the train_val, and test cohorts.

Table 1 Patient characteristics

| Characteristics | All | Train_Val | Test |
|---|---|---|---|
| Number of Patients | 331 | 265 | 66 |
| Age (median, range) | 65 (28-89) | 65 (29-89) | 65 (28-87) |
| Gender (male) | 201 (60.7%) | 162 (61.1%) | 39 (59.1%) |

| Type of Rectal Cancer | | | |
|---|---|---|---|
| EVI (+) | 103 (31.1%) | 83 (31.3%) | 20 (30.3%) |
| MFI (+) | 78 (23.6%) | 63 (23.8%) | 15 (22.7%) |
| Medical Center | | | |
| La Fe | 194 (58.6%) | 158 (59.6%) | 36 (54.5%) |
| ULS | 113 (34.1%) | 87 (32.8%) | 26 (39.4%) |
| CHU Angers | 24 (7.3%) | 20 (7.5%) | 4 (6.1%) |
| Manufacturer | | | |
| GE | 221 (66.8%) | 183 (69.1%) | 38 (57.6%) |
| Siemens | 106 (32.0%) | 80 (30.2%) | 26 (39.4%) |
| Other | 4 (1.2%) | 2 (0.7%) | 2 (3.0%) |

Values for age are median (range); all other values are counts with percentages in parentheses. La Fe = La Fe University and Polytechnic Hospital; ULS = Sant'Andrea University Hospital, Sapienza University of Rome; CHU Angers = Centre Hospitalier Universitaire d'Angers. EVI = extramural vascular invasion; MFI = mesorectal fascia invasion; "+" indicates presence of the feature.

## Classification results

We evaluated SeResNet, UMedPT, and UMedPT_LR for classifying EVI and MFI on axial and sagittal T2-weighted MRI, with and without harmonization. All performance metrics reported below were obtained on the test set. Overall, our models consistently outperformed the CHAIMELEON challenge winner's model[28], demonstrating superior generalizability across orientations and preprocessing conditions.

For EVI classification, UMedPT_LR achieved the strongest performance. On single-plane patches, it obtained an AUC of 0.76 (balanced accuracy 0.70) on axial images and 0.73 (balanced accuracy 0.68) on sagittal images, surpassing both SeResNet and UMedPT (**Table 2**). When axial and sagittal planes were fused, the UMedPT_LR model achieved the highest performance with an AUC of 0.82, sensitivity of 0.75, and F1 score of 0.73 (**Table 4**). As illustrated in the ROC curves (**Fig.5.**, left), the fused model provided the best discrimination among all evaluated approaches and outperformed the CHAIMELEON challenge winner (AUC = 0.74). Notably, harmonization impaired EVI detection, with performance decreasing substantially across both orientations.

For MFI classification, UMedPT demonstrated the best performance on axial harmonized patches, achieving an AUC of 0.77 and balanced accuracy of 0.72, representing the highest MFI result across all conditions (**Table 3**) and exceeding the CHAIMELEON challenge winner (AUC = 0.75). UMedPT_LR performed comparably on axial original patches (AUC 0.75, balanced accuracy 0.71; **Table 3**), while sagittal results were uniformly lower, with modest gains from SeResNet on harmonized sagittal images (AUC 0.73; **Table 3**). Multiplanar fusion of axial and sagittal patches using UMedPT_LR yielded an AUC of 0.71, which did not surpass the best single-plane performance. This pattern is also evident in the ROC curves (**Fig.4.**, right), where UMedPT on harmonized axial images achieves the best performance.

Table 2 Diagnostic performance of different models for classifying **EVI** on axial and sagittal T2-weighted MRI, before and after harmonization (test set)

| **Plane** | **Model** | **AUC** (95% CI) | **Sensitivity** (95% CI) | **Specificity** (95% CI) | **F1** (95% CI) | **Balanced_acc** (95% CI) |
|---|---|---|---|---|---|---|
| Axial Original | SeResnet | 0.57 (0.41-0.71) | 0.70 (0.47-0.89) | 0.44 (0.31-0.59) | 0.48 (0.30-0.63) | 0.57 (0.44-0.70) |
| | UMedPT | 0.73 (0.57-0.86) | 0.65 (0.43-0.85) | 0.67 (0.53-0.81) | 0.54 (0.35-0.70) | 0.66 (0.53-0.78) |
| | UMedPT_LR | **0.76 (0.62-0.89)** | 0.60 (0.35-0.82) | 0.80 (0.68-0.91) | 0.59 (0.36-0.74) | 0.70 (0.58-0.82) |
| Axial Harmonized | SeResnet | 0.60 (0.45-0.75) | 0.85 (0.67-1.0) | 0.36 (0.22-0.50) | 0.52 (0.36-0.65) | 0.60 (0.49-0.70) |
| | UMedPT | **0.65 (0.49-0.80)** | 0.60 (0.37-0.80) | 0.64 (0.51-0.79) | 0.50 (0.32-0.67) | 0.62 (0.49-0.75) |
| | UMedPT_LR | 0.64 (0.47-0.78) | 0.45 (0.22-0.67) | 0.71 (0.58-0.84) | 0.43 (0.21-0.61) | 0.58 (0.45-0.71) |
| Sagittal Original | SeResnet | 0.52 (0.35-0.68) | 0.55 (0.31-0.75) | 0.58 (0.43-0.72) | 0.44 (0.25-0.59) | 0.56 (0.42-0.69) |
| | UMedPT | 0.67 (0.53-0.82) | 0.50 (0.28-0.74) | 0.71 (0.58-0.84) | 0.47 (0.26-0.64) | 0.61 (0.47-0.74) |
| | UMedPT_LR | **0.73 (0.60-0.85)** | 0.60 (0.38-0.81) | 0.76 (0.63-0.87) | 0.56 (0.35-0.73) | 0.68 (0.55-0.80) |
| Sagittal Harmonized | SeResnet | **0.63 (0.47-0.79)** | 0.70 (0.49-0.91) | 0.53 (0.39-0.68) | 0.51 (0.33-0.67) | 0.62 (0.49-0.74) |
| | UMedPT | 0.53 (0.36-0.69) | 0.0 (0.0-0.0) | 1.0 (1.0-1.0) | 0.0 (0.0-0.0) | 0.50 (0.50-0.50) |
| | UMedPT_LR | 0.55 (0.38-0.72) | 0.35 (0.15-0.57) | 0.80 (0.69-0.91) | 0.39 (0.18-0.57) | 0.58 (0.45-0.69) |

Table 3 Diagnostic performance of different models for classifying **MFI** on axial and sagittal T2-weighted MRI, before and after harmonization (test set)

| **Plane** | **Model** | **AUC** (95% CI) | **Sensitivity** (95% CI) | **Specificity** (95% CI) | **F1** (95% CI) | **Balanced_acc** (95% CI) |
|---|---|---|---|---|---|---|
| Axial Original | SeResnet | 0.75 (0.61-0.87) | 0.73 (0.50-0.93) | 0.64 (0.50-0.78) | 0.50 (0.30 - 0.67) | 0.69 (0.55-0.81) |
| | UMedPT | 0.70 (0.55-0.84) | 0.60 (0.36-0.83) | 0.66 (0.53-0.78) | 0.44 (0.24 - 0.62) | 0.63 (0.50-0.78) |
| | UMedPT_LR | **0.75 (0.60-0.89)** | 0.60 (0.33-0.85) | 0.82 (0.71-0.93) | 0.55 (0.31 - 0.74) | 0.71 (0.57-0.85) |
| Axial Harmonized | SeResnet | 0.74 (0.58-0.87) | 0.60 (0.35-0.84) | 0.70 (0.57-0.82) | 0.46 (0.26 - 0.65) | 0.65 (0.52-0.79) |
| | UMedPT | **0.77 (0.65-0.89)** | 0.73 (0.50-0.94) | 0.70 (0.57-0.82) | 0.54 (0.34 - 0.71) | 0.72 (0.59-0.85) |
| | UMedPT_LR | 0.72 (0.59-0.86) | 0.40 (0.14-0.67) | 0.90 (0.81-0.97) | 0.46 (0.19 - 0.68) | 0.65 (0.52-0.79) |
| Sagittal Original | SeResnet | 0.41 (0.27-0.59) | 0.87 (0.67-1.0) | 0.14 (0.06-0.24) | 0.37 (0.22 - 0.50) | 0.50 (0.39-0.60) |
| | UMedPT | 0.61 (0.45-0.78) | 0.53 (0.26-0.79) | 0.66 (0.53-0.80) | 0.40 (0.19 - 0.59) | 0.60 (0.45-0.74) |
| | UMedPT_LR | **0.62 (0.46-0.78)** | 0.40 (0.14-0.67) | 0.80 (0.68-0.91) | 0.39 (0.16 - 0.59) | 0.60 (0.47-0.74) |
| Sagittal Harmonized | SeResnet | **0.73 (0.56-0.89)** | 0.73 (0.50-0.95) | 0.70 (0.57-0.83) | 0.54 (0.33 - 0.71) | 0.72 (0.58-0.85) |
| | UMedPT | 0.59 (0.44-0.75) | 0.53 (0.29-0.80) | 0.62 (0.48-0.76) | 0.38 (0.18 - 0.56) | 0.58 (0.43-0.72) |
| | UMedPT_ LR | 0.60 (0.46-0.76) | 0.27 (0.07-0.50) | 0.76 (0.64-0.88) | 0.26 (0.07 - 0.46) | 0.51 (0.40-0.65) |

Table 4 Diagnostic performance of multiplanar fusion for UMedPT_LR model on EVI and MFI classification (test set)

| **Multiplanar Fusion** | **AUC** (EVI) | **AUC** (MFI) | **Sensitivity** (EVI) | **Sensitivity** (MFI) | **F1** (EVI) | **F1** (MFI) |
|---|---|---|---|---|---|---|
| Axial Original + Sagittal Original | 0.82 | 0.71 | 0.75 | 0.73 | 0.73 | 0.51 |
| Axial Harmonized + Sagittal Harmonized | 0.58 | 0.75 | 0.35 | 0.07 | 0.37 | 0.13 |

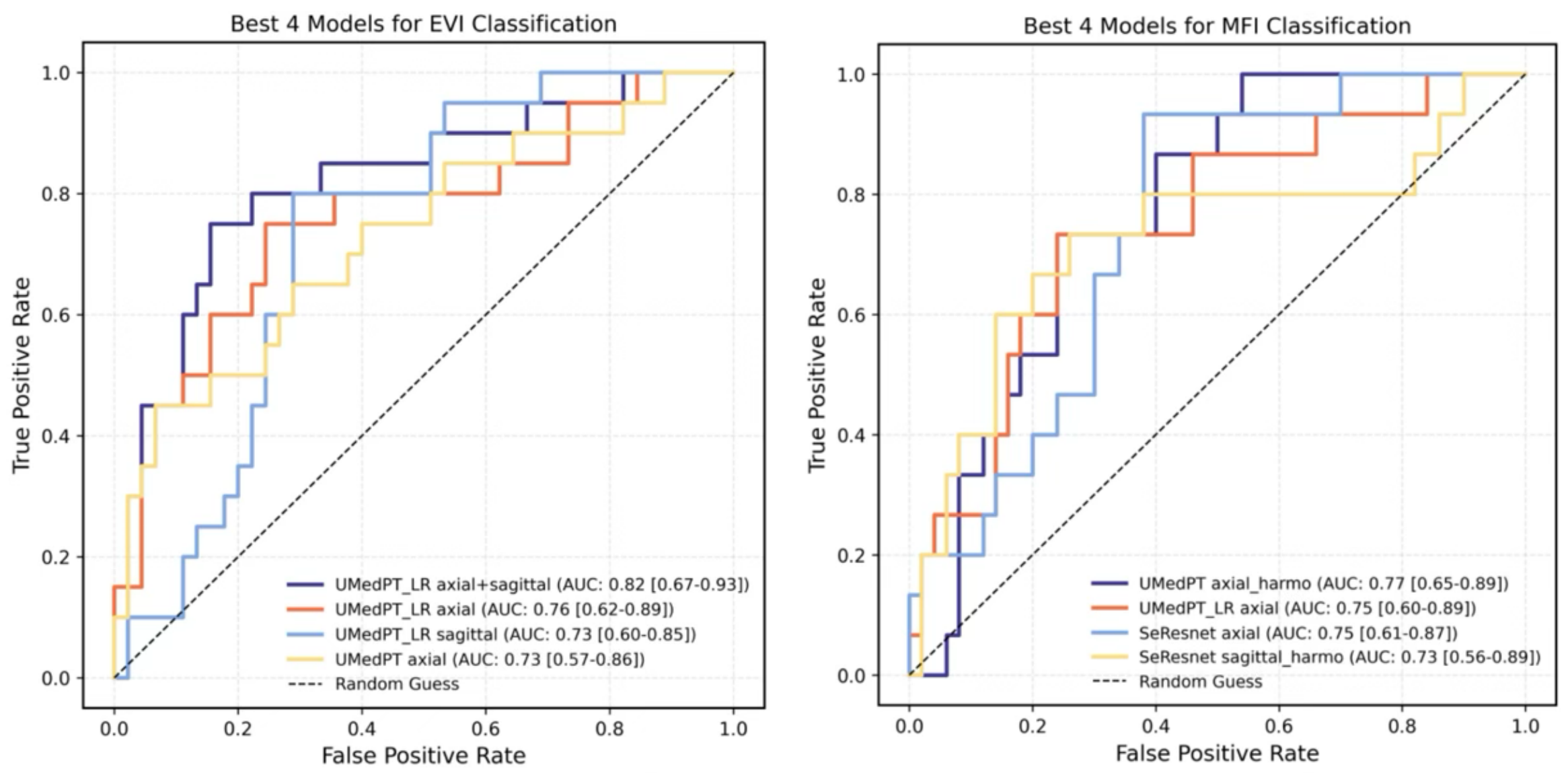

Fig.4. ROC curves of the top four models for classifying extramural vascular invasion (EVI, left) and mesorectal fascia invasion (MFI, right) on the CHAIMELEON's testset. For EVI, UMedPT_LR with multiplanar fusion (axial + sagittal planes) achieved the highest AUC = 0.82. For MFI, the best performance was UMedPT trained on axial harmonized patches (AUC = 0.77). "Axial" refers to the original axial patch, and "axial_harmo" indicates the axial harmonized patch.

## Grad-CAM visualization

To enhance interpretability, we applied Grad-CAM on axial T2-weighted MR images to highlight regions that most influenced the model's predictions. As shown in **Fig.5.**, each pair displays the original image and its corresponding heatmap, highlighting areas of model attention.

For EVI classification, the model predominantly focused on localized areas near the tumor boundary. In contrast, MFI predictions consistently involved broader regions, particularly around the mesorectal fascia.

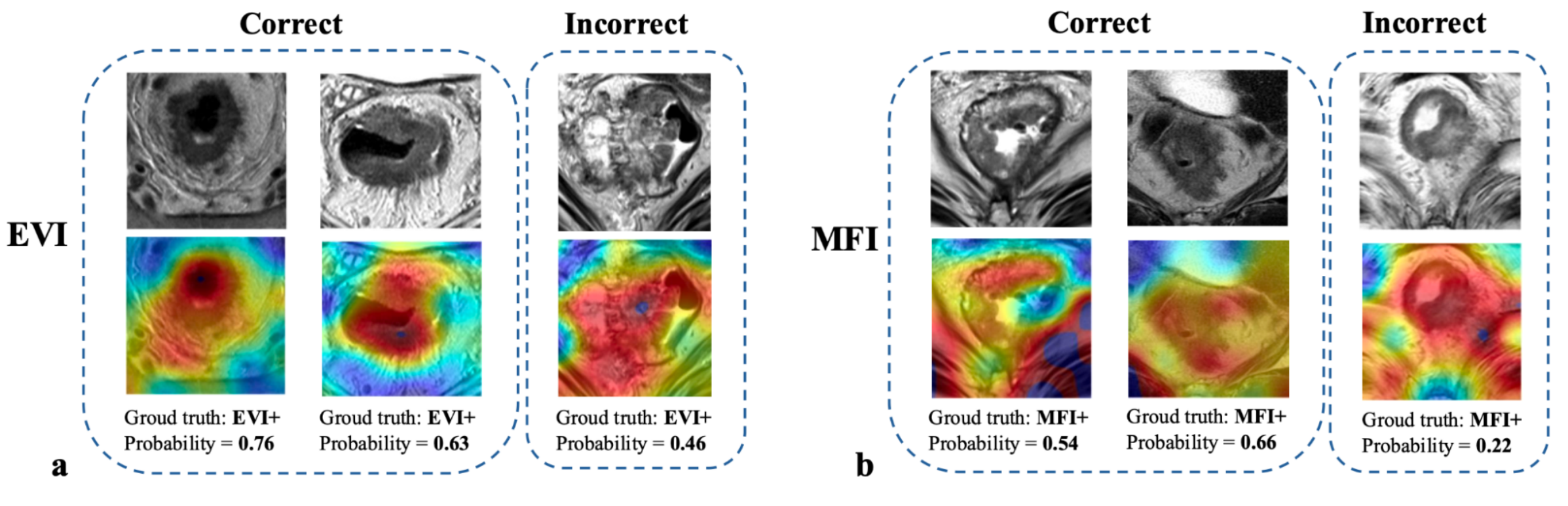

Fig. 5. Grad-CAM visualizations for EVI and MFI classification. Examples of axial T2-weighted MR images (top row in each case) and corresponding Grad-CAM heatmaps (bottom row) are shown for extramural vascular invasion (EVI, a) and mesorectal fascia invasion (MFI, b). For each condition, two correctly classified cases (left) and one misclassified case (right) are presented. Grad-CAM heatmaps highlight image regions that contributed most to the model's prediction. Ground truth labels and predicted probabilities are displayed beneath each case.

### RQS score

The proposed framework achieved an RQS 2.0 score of 25 out of 35 (71%), corresponding to Radiomics Readiness Level 6, which indicates a validated multi-center study demonstrating good reproducibility and potential for clinical translation. Detailed category-wise scores and figures are presented in the **Appendix A4.**

## Discussion

In this multi-centre cohort, foundation-model pipelines (UMedPT/UMedPT_LR) outperformed conventional CNNs for both EVI and MFI classification tasks. For EVI, multiplanar UMedPT_LR achieved an AUC of 0.82 (vs SeResNet 0.57–0.60), indicating the value of pretrained representations and cross-plane feature aggregation. For MFI, UMedPT on axial harmonized patches reached 0.77. These findings position foundation models, multiplanar fusion, and harmonization as complementary strategies for robust rectal MRI analysis.

Previous radiomics-based studies have demonstrated promising results in rectal cancer classification. For EVI, Shu[33] et al. and Liu[34] et al. both reported improved AUC ( $\geq 0.83$) when combining handcrafted radiomics features with clinical variables in integrated nomograms. For MFI, Liang[35] *et al.* (n=301) achieved an AUC of 0.91 using a combined model incorporating tumor and fascia region radiomics. However, these approaches typically depend on handcrafted, site-specific feature engineering, which may hinder generalizability and scalability. A recent deep learning study by Cai[36] et al. developed an automated pipeline combining nnU-Net segmentation with an MLNet classifier and achieved an external AUC of 0.76 for EVI classification using DWI across nine centres. Our foundation model–based framework achieved higher EVI performance (AUC=0.82) despite being fine-tuned on a smaller three-centre dataset (n=331), highlighting the superior generalizability and data efficiency of foundation model representations over task-specific CNNs.

Most prior studies on rectal cancer MRI local staging have focused on axial T2-weighted images, with sagittal images mainly used for tumor localization or assessment of the anal canal and pelvic floor[7, 12]. To our knowledge, this is one of the first studies to quantitatively compare axial and sagittal T2-weighted images and multiplanar fusion for automated classification of EVI and MFI. In our experiments, axial plane consistently yield higher performance for both tasks, likely due to their broader coverage of vascular and fascial structures. Multiplanar fusion improved performance when complementary information was present across planes[37], though gains were limited when one orientation dominated. This may highlight the importance of explicit plane selection in rectal MRI AI applications.

We also observed that harmonization improved MFI classification, particularly for UMedPT on axial images, but reduced performance for EVI. This suggests that excessive harmonization may attenuate subtle diagnostic cues[38, 39], especially those reflecting vascular abnormalities. These results indicate that harmonization may not be universally beneficial and should be adapted to the underlying biological question and the model’s feature dependencies[40].

This study has several limitations. Although data were obtained from three centers, we did not analyze center-specific effects in detail. The overall sample size was modest (n=331), with only 66 cases in the test set, which may limit statistical power and generalizability. While the strong feature representation of foundation models may have mitigated some of these effects[41, 42], further validation in larger, multi-institutional cohorts is warranted. Future research should include prospective validation across broader populations, integration of diffusion-weighted and dynamic contrast-enhanced sequences, and longitudinal outcome prediction to support individual treatment decisions. Moreover, combining imaging biomarkers with molecular and genomic data could further enhance predictive accuracy and advance precision oncology. Finally, *in silico* trials comparing diagnostic performance across radiologists alone, AI alone, and AI-assisted radiologists would provide valuable evidence of clinical translation, as demonstrated in our previous studies (Sergei et al[43]; Xian et al[44]).

## Conclusion

In summary, the Foundation models UMedPT and UMedPT_LR achieved state-of-the-art performance for EVI and MFI classification from multi-center rectal MRI. Multiplanar fusion (axial + sagittal planes), frequency domain harmonization, and consistent center-cropped rectal patches contributed to performance gains. Our findings highlight the potential of foundation models in rectal cancer staging and support their future application in clinical workflows.

### Abbreviations

EVI Extramural vascular invasion

MFI Mesorectal fascia invasion

CNNs Convolutional neural networks

UMedPT The universal biomedical pretrained transformer

UMedPT_LR A logistic-regression variant using frozen UMedPT features

MLP Multilayer perceptron

Grad-CAM Gradient-weighted class activation mapping

TME Total mesorectal excision

nCRT Neoadjuvant chemoradiotherapy

CRM Circumferential resection margin

AI Artificial intelligence

FFT Fast fourier transform

SeResNet Squeeze and excitation residual network

PCA Principal component analysis

CI Confidence intervals

ROC Receiver operating characteristic

AUC Area under the curve

## Declarations

### Ethical Approval

This study was conducted using fully anonymized data provided by the CHAIMELEON project, which adheres to the EU General Data Protection Regulation. As the dataset is publicly available for research purposes, additional ethical approval and informed consent were not required.

### Consent for Publication

Not applicable.

### Data and Code Availability

The multicentre MRI datasets used in this study are subject to institutional and EU data protection regulations and therefore cannot be publicly shared. The code will be made publicly available at https://github.com/yumengzhang97/foundation-model-rectal-mri upon publication.

### Grants and Funding

Authors acknowledge financial support from ERC advanced grant (ERC-ADG-2015 n° 694812 - Hypoximmuno),, ERC-2020-PoC: 957565-AUTO.DISTINCT. Authors also acknowledge financial support from the European Union's Horizon research and innovation programme under grant agreement: CHAIMELEON n° 952172 (main contributor), ImmunoSABR n° 733008, EuCanImage n° 952103, TRANSCAN Joint Transnational Call 2016 (JTC2016 CLEARLY n° UM 2017-8295), IMI-OPTIMA n° 101034347, AIDAVA (HORIZON-HLTH-2021-TOOL-06) n°101057062, REALM (HORIZON-HLTH-2022-TOOL-11) n° 101095435, RADIOVAL (HORIZON-HLTH-2021-DISEASE-04-04) n°101057699 and EUCAIM (DIGITAL-2022-CLOUD-AI-02) n°101100633. This study was also supported by the China Scholarship Council grant (202208110055).

### Disclosures

Disclosures from the last 36 months within and outside the submitted work: none related to the current manuscript; outside of current manuscript: grants/sponsored research agreements from Radiomics SA, Convert Pharmaceuticals and LivingMed Biotech. He received a presenter fee (in cash or in kind) and/or reimbursement of travel costs/consultancy fee (in cash or in kind) from Radiomics SA, BHV & Roche. PL has shares in the companies Radiomics SA, Convert pharmaceuticals, Comunicare, LivingMed Biotech, BHV and Bactam. PL is co-inventor of two issued patents with royalties on radiomics (PCT/NL2014/050248 and PCT/NL2014/050728), licensed to Radiomics SA; one issued patent on mtDNA (PCT/EP2014/059089), licensed to ptTheragnostic/DNAmito; one non-issued patent on LSRT (PCT/ P126537PC00, US: 17802766), licensed to Varian; three non-patented inventions (softwares) licensed to ptTheragnostic/DNAmito, Radiomics SA and Health Innovation Ventures and two non-issued, non-licensed patents on Deep Learning-Radiomics (N2024482, N2024889). He confirms that none of the above entities were involved in the preparation of this paper.

## Authors' Contributions

Conceptualization, Z.S., P.L., and L.M.-B.; methodology, Y.Z., S.A.M., D.K. and Z.S.; formal analysis, Y.Z., S.A.M., and Z.S.; data curation, Y.Z., Z.S., G.R., S.F.-A., M.Z. and C.A.; writing—original draft preparation, Y.Z. and S.A.M.; writing—review and editing, Y.Z., S.A.M., S.A., E.I.-C., A.J.-P., L.M.-B., Z.S. and P.L.; visualization, Y.Z., S.A.M. and Z.S.; supervision, Z.S. and P.L.; funding acquisition, L.M.-B. and P.L. All authors have read and agreed to the published version of the manuscript.

# References

1. Bray F, Laversanne M, Sung H, et al (2024) Global cancer statistics 2022: GLOBOCAN estimates of incidence and mortality worldwide for 36 cancers in 185 countries. CA Cancer J Clin 74:229–263

2. Keller DS, Berho M, Perez RO, et al (2020) The multidisciplinary management of rectal cancer. Nat Rev Gastroenterol Hepatol 17:414–429

3. Morgan E, Arnold M, Gini A, et al (2023) Global burden of colorectal cancer in 2020 and 2040: incidence and mortality estimates from GLOBOCAN. Gut 72:338–344

4. Emile SH, Horesh N, Garoufalia Z, et al (2024) Accuracy of clinical staging of localized colon cancer: A National Cancer Database cohort analysis. Ann Surg Oncol 31:6461–6469

5. Siddiqui MRS, Simillis C, Hunter C, et al (2017) A meta-analysis comparing the risk of metastases in patients with rectal cancer and MRI-detected extramural vascular invasion (mrEMVI) vs mrEMVI-negative cases. British Journal of Cancer 116:1513–1519

6. Hofheinz R-D, Fokas E, Benhaim L, et al (2025) Localised rectal cancer: ESMO Clinical Practice Guideline for diagnosis, treatment and follow-up. Ann Oncol 36:1007–1024

7. Horvat N, Carlos Tavares Rocha C, Clemente Oliveira B, et al (2019) MRI of rectal cancer: Tumor staging, imaging techniques, and management. Radiographics 39:367–387

8. Inoue A, Sheedy SP, Heiken JP, et al (2021) MRI-detected extramural venous invasion of rectal cancer: Multimodality performance and implications at baseline imaging and after neoadjuvant therapy. Insights Imaging 12:110

9. Sohn B, Lim J-S, Kim H, et al (2015) MRI-detected extramural vascular invasion is an independent prognostic factor for synchronous metastasis in patients with rectal cancer. Eur Radiol 25:1347–1355

10. Gu C, Yang X, Zhang X, et al (2019) The prognostic significance of MRI-detected extramural venous invasion, mesorectal extension, and lymph node status in clinical T3 mid-low rectal cancer. Sci Rep 9:12523

11. Santiago I, Figueiredo N, Parés O, Matos C (2020) MRI of rectal cancer-relevant anatomy and staging key points. Insights Imaging 11:100

12. Bogveradze N, Snaebjornsson P, Grotenhuis BA, et al (2023) MRI anatomy of the rectum: key concepts important for rectal cancer staging and treatment planning. Insights Imaging 14:13

13. Zhang X-Y, Wang S, Li X-T, et al (2018) MRI of extramural venous invasion in locally advanced rectal cancer: Relationship to tumor recurrence and overall survival. Radiology 289:677–685

14. Lambin P, Leijenaar RTH, Deist TM, et al (2017) Radiomics: the bridge between medical imaging and personalized medicine. Nat Rev Clin Oncol 14:749–762

15. Lambin P, Rios-Velazquez E, Leijenaar R, et al (2012) Radiomics: extracting more information from medical images using advanced feature analysis. Eur J Cancer 48:441–446

16. Aerts HJWL, Velazquez ER, Leijenaar RTH, et al (2014) Decoding tumour phenotype by noninvasive imaging using a quantitative radiomics approach. Nat Commun 5:4006

17. Jiang X, Zhao H, Saldanha OL, et al (2023) An MRI Deep learning model predicts outcome in rectal cancer. Radiology 307:e222223

18. Mali SA, Ibrahim A, Woodruff HC, et al (2021) Making radiomics more reproducible across scanner and imaging protocol variations: A review of harmonization methods. J Pers Med 11:842

19. Schurink NW, van Kranen SR, Roberti S, et al (2022) Sources of variation in multicenter rectal MRI data and their effect on radiomics feature reproducibility. Eur Radiol 32:1506–1516

20. Wichtmann BD, Albert S, Zhao W, et al (2022) Are we there yet? The value of deep learning in a multicenter setting for response prediction of locally advanced rectal cancer to neoadjuvant chemoradiotherapy. Diagnostics (Basel) 12:1601

21. Pomponio R, Erus G, Habes M, et al (2020) Harmonization of large MRI datasets for the analysis of brain imaging patterns throughout the lifespan. Neuroimage 208:116450

22. Modanwal G, Vellal A, Buda M, Mazurowski MA (2020) MRI image harmonization using cycle-consistent generative adversarial network. In: Hahn HK, Mazurowski MA (eds) Medical Imaging 2020: Computer-Aided Diagnosis. SPIE

23. Tian C, Ma X, Lu H, et al (2023) Deep learning models for preoperative T-stage assessment in rectal cancer using MRI: exploring the impact of rectal filling. Front Med (Lausanne) 10:1326324

24. Azad B, Azad R, Eskandari S, et al (2023) Foundational models in medical imaging: A comprehensive survey and future vision. arXiv [cs.CV]

25. Engelmann J, Bernabeu MO (2025) Training a high-performance retinal foundation model with half-the-data and 400 times less compute. Nat Commun 16:6862

26. Yang Z, Wei T, Liang Y, et al (2025) A foundation model for generalizable cancer diagnosis and survival prediction from histopathological images. Nat Commun 16:2366

27. Schäfer R, Nicke T, Höfener H, et al (2024) Overcoming data scarcity in biomedical imaging with a foundational multi-task model. Nat Comput Sci 4:495–509

28. CHAIMELEON Open Challenges - Grand Challenge. In: grand-challenge.org. https://chaimeleon.grand-challenge.org/evaluation/e0dc2f29-f8ba-4184-948f-327cc6877982/. Accessed 23 Apr 2025

29. Akinci D'Antonoli T, Berger LK, Indrakanti AK, et al (2025) TotalSegmentator MRI: Robust sequence-independent segmentation of multiple anatomic structures in MRI. Radiology 314:e241613

30. Cardoso MJ, Li W, Brown R, et al (2022) MONAI: An open-source framework for deep learning in healthcare. arXiv [cs.LG]

31. Selvaraju RR, Cogswell M, Das A, et al (2020) Grad-CAM: Visual explanations from deep networks via gradient-based localization. Int J Comput Vis 128:336–359

32. Lambin P, Woodruff HC, Mali SA, et al (2025) Radiomics Quality Score 2.0: towards radiomics readiness levels and clinical translation for personalized medicine. Nat Rev Clin Oncol. https://doi.org/10.1038/s41571-025-01067-1

33. Shu Z, Mao D, Song Q, et al (2022) Multiparameter MRI-based radiomics for preoperative prediction of extramural venous invasion in rectal cancer. Eur Radiol 32:1002–1013

34. Liu S, Yu X, Yang S, et al (2021) Machine learning-based radiomics nomogram for detecting extramural venous invasion in rectal cancer. Front Oncol 11:610338

35. Liang H, Ma D, Ma Y, et al (2024) Comparison of conventional MRI analysis versus MRI-based radiomics to predict the circumferential margin resection involvement of rectal cancer. BMC Gastroenterol 24:209

36. Cai L, Lambregts DMJ, Beets GL, et al (2024) An automated deep learning pipeline for EMVI classification and response prediction of rectal cancer using baseline MRI: a multi-centre study. NPJ Precis Oncol 8:17

37. Zhang Q, Long Y, Cai H, et al (2024) A multi-slice attention fusion and multi-view personalized fusion lightweight network for Alzheimer's disease diagnosis. BMC Med Imaging 24:258

38. Orlhac F, Boughdad S, Philippe C, et al (2018) A postreconstruction harmonization method for multicenter radiomic studies in PET. J Nucl Med 59:1321–1328

39. Ibrahim A, Primakov S, Beuque M, et al (2021) Radiomics for precision medicine: Current challenges, future prospects, and the proposal of a new framework. Methods 188:20–29

40. An L, Chen J, Chen P, et al (2022) Goal-specific brain MRI harmonization. Neuroimage 263:119570

41. Moor M, Banerjee O, Abad ZSH, et al (2023) Foundation models for generalist medical artificial intelligence. Nature 616:259–265

42. Zhou Y, Chia MA, Wagner SK, et al (2023) A foundation model for generalizable disease detection from retinal images. Nature 622:156–163

43. Primakov SP, Ibrahim A, van Timmeren JE, et al (2022) Automated detection and segmentation of non-small cell lung cancer computed tomography images. Nat Commun 13.: https://doi.org/10.1038/s41467-022-30841-3

44. Zhong X, Salahuddin Z, Chen Y, et al (2024) Methodological explainability evaluation of an interpretable deep learning model for post-hepatectomy liver failure prediction incorporating counterfactual explanations and layerwise relevance propagation: A prospective in silico trial. arXiv [cs.CV]

# Appendix

### A1. Frequency domain harmonization methods

This method is designed to reduce scanner and protocol induced intensity differences while preserving clinically relevant anatomical features essential for tumor staging and invasion assessment. A curated reference set of 17 high-quality rectal T2w MRI cases was selected based on consistent acquisition parameters (pulse sequence, TE, TR) and optimal image quality, as assessed by an expert radiologist. These reference scans were used to define the harmonization target space, representing the diversity of tissue-contrast distributions observed across sites. Synthetic contrast perturbations were applied to these reference images by modifying their frequency domain representations via Fast Fourier Transform (FFT). A circular mask centered in the frequency matrix was used to iteratively vary spatial frequency content, simulating a range of inter-site contrast conditions. Each altered image was then reconstructed into the spatial domain using inverse FFT, yielding 300 synthetic variants per 2D slice.

A self-supervised convolutional autoencoder was trained to reconstruct harmonized images from these contrast-altered inputs. The method is considered self-supervised because it synthetically perturbed images as inputs and the original reference images as targets, eliminating the need for manual labels. The architecture was based on a modified U-Net3+[1] with a VGG16[2] encoder pretrained on ImageNet, enhanced by learnable convolutional pooling and skip connections. Training was performed in two stages: (i) freezing the encoder while training decoder layers on 40,000 synthetic images followed by (ii) fine-tuning the entire model on additional 50,320 images for 200 epochs. The optimization used a composite loss combining mean squared error and L2 regularization to ensure fidelity in image reconstruction. This harmonization model produces standardized contrast outputs from heterogeneous MRI inputs and its outputs were included in our pipeline to improve robustness in downstream classification tasks.

### A2. SeResNet baseline

We employed a modified version of Squeeze and Excitation Residual Network (SeResNet)-34 as the backbone network for classification, tailored to the anisotropic spatial resolution typical in pelvic MRI. To preserve spatial detail while minimizing feature loss through early downsampling, we customized the network stem to use a convolutional kernel of size 3×3×1 and applied anisotropic strides of (2,2,1) in the first two stages. LeakyReLU was used in place of ReLU to improve gradient flow in low-contrast regions.

The model comprises five sequential ResidualEncoder stages, with output channel sizes of 32, 64, 128, 256, and 512, respectively. The convolutional kernel size for the first two stages are 3×3×1 and 3×3×3 for the remaining ones. The stride configuration is (2,2,1) for the first two stages and the final stages, followed by (2,2,2) for the remaining stages. Each stage consists of 1, 3, 4, 6, and 3 residual blocks[3], respectively. All residual units are enhanced with squeeze-and-excitation (SE)[4] modules, enabling adaptive channel-wise feature recalibration. Following the final stage, a 3D global average pooling layer compresses the spatial dimensions, and the resulting feature vector is passed to a fully connected layer with 512 units. A final sigmoid-activated output layer produces binary predictions for EVI or MFI classification. The full architecture is summarized in **Figure A1.**

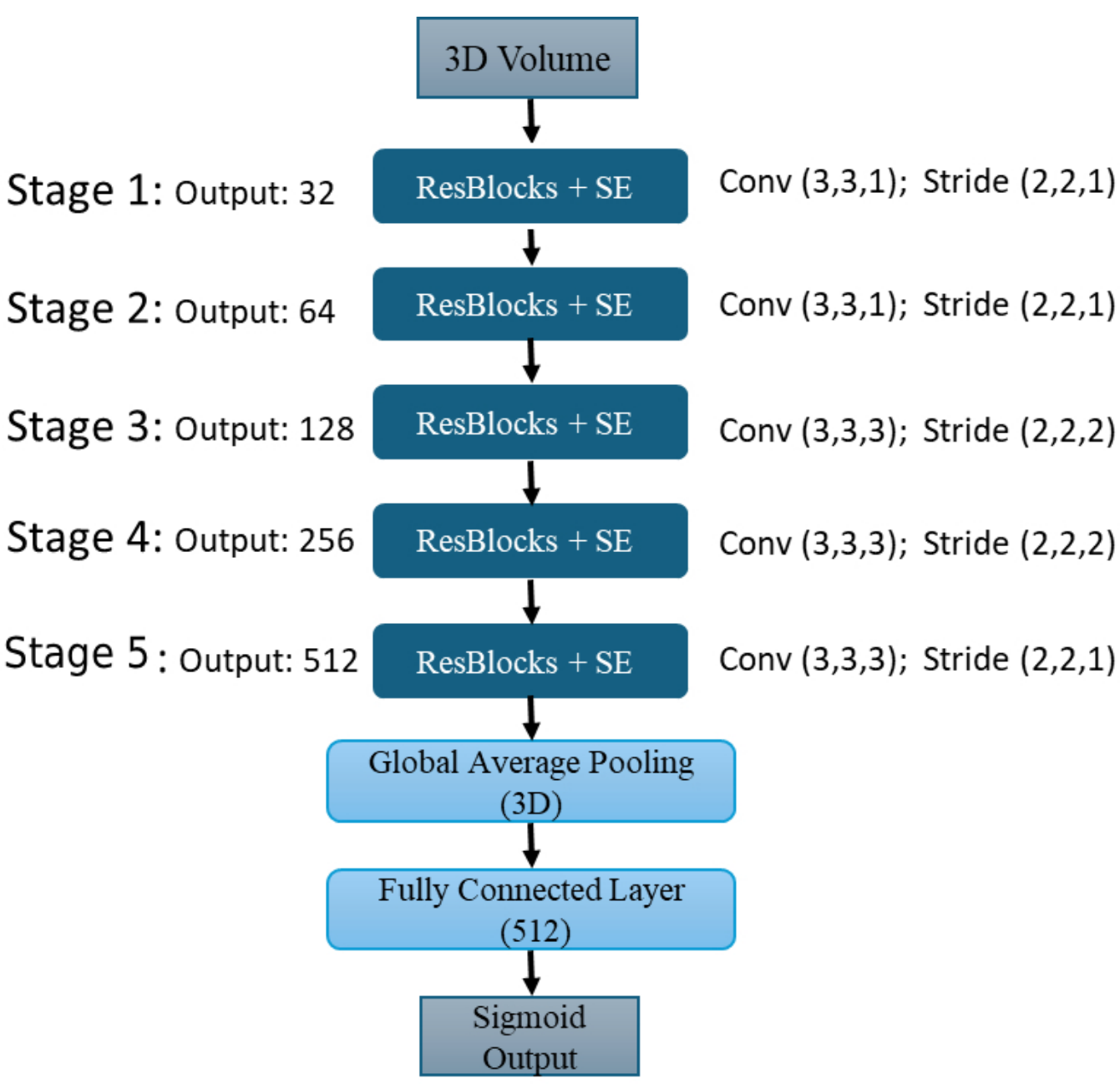

**Figure A1.** Architecture of the modified SeResNet used for EVI and MFI classification. The model consists of five convolutional stages, each comprising residual blocks enhanced with SE modules. Anisotropic convolution kernels and strides (e.g., 3×3×1 with stride 2×2×1) are used in the early stages to accommodate the non-isotropic resolution of pelvic MRI. The final feature map is globally pooled and passed through a 512-unit fully connected layer, followed by a sigmoid output for binary classification. Channel dimensions are progressively expanded across stages, as shown on the left.

### A3. Data augmentation parameters

**Table A1** The parameters of data augmentation(applied with MONAI)

| Transform | Probability | Parameters |
|---|---|---|
| RandZoomd | 0.2 | min_zoom=0.9, max_zoom=1.1 |
| RandAffined | 0.5 | rotate_range=rotate_range=(0, 0, π/15), shear_range=(0.1, 0.1, 0.1), scale_range=(0.1, 0.1, 0.1), padding_mode='border' |
| RandFlipd | 0.2 | spatial_axias=0 |
| RandGaussianNoised | 0.1 | mean=0.0, std=0.1 |
| RandGaussianSmoothd | 0.1 | sigma_x=(0.5, 1), sigma_y=(0.5, 1), sigma_z=(0.5, 1) |
| RandScaleIntensityd | 0.2 | factors=(0.8, 1.2) |
| RandAdjustContrastd | 0.2 | gamma=(0.8, 1.2) |
| RandBiasFieldd | 0.1 | coeff_range=(0.1, 0.2) |
| RandGibbsNoised | 0.1 | alpha=(0.6, 0.8) |

### A4. Radiomics Quality Score 2.0

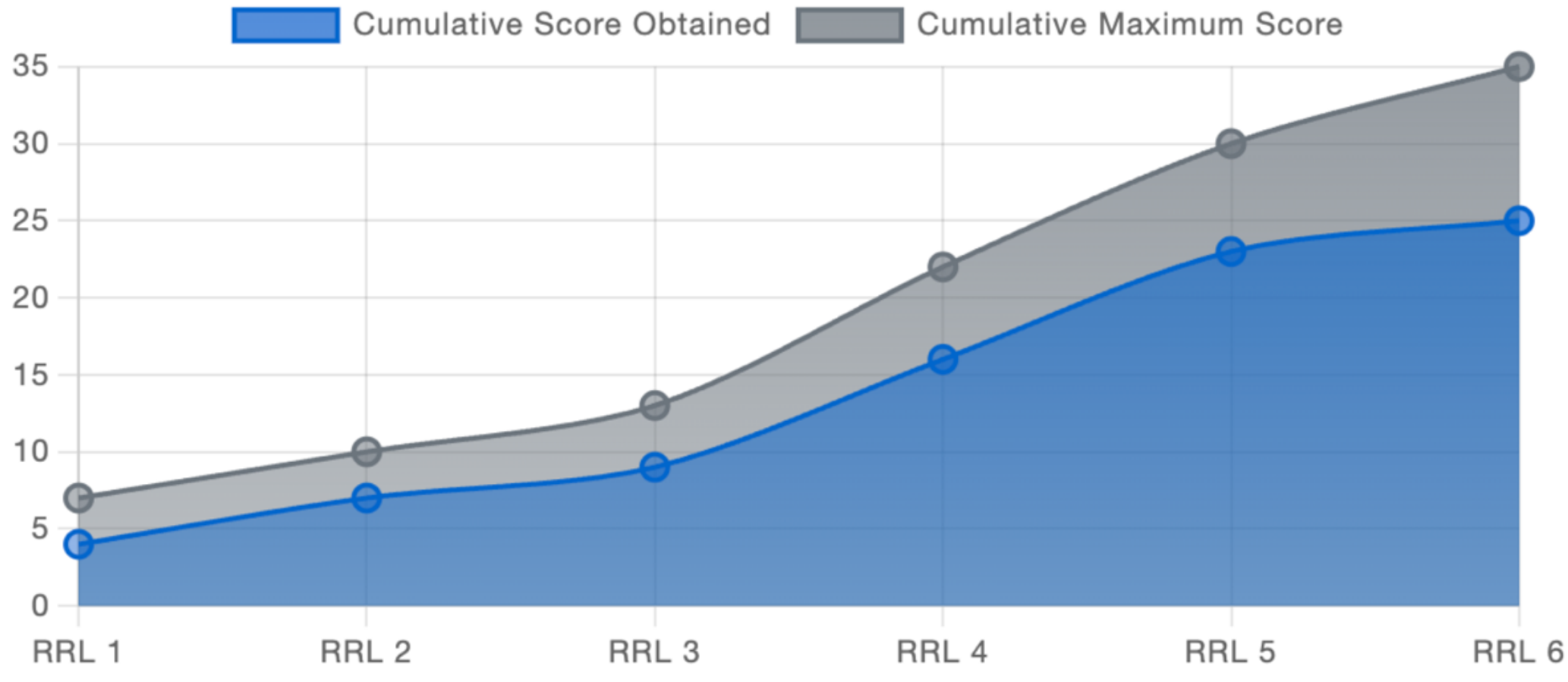

**Figure A2.** Evolution of score by RRL level. Cumulative score obtained (blue) versus cumulative maximum score (grey) across RRL1–RRL6.

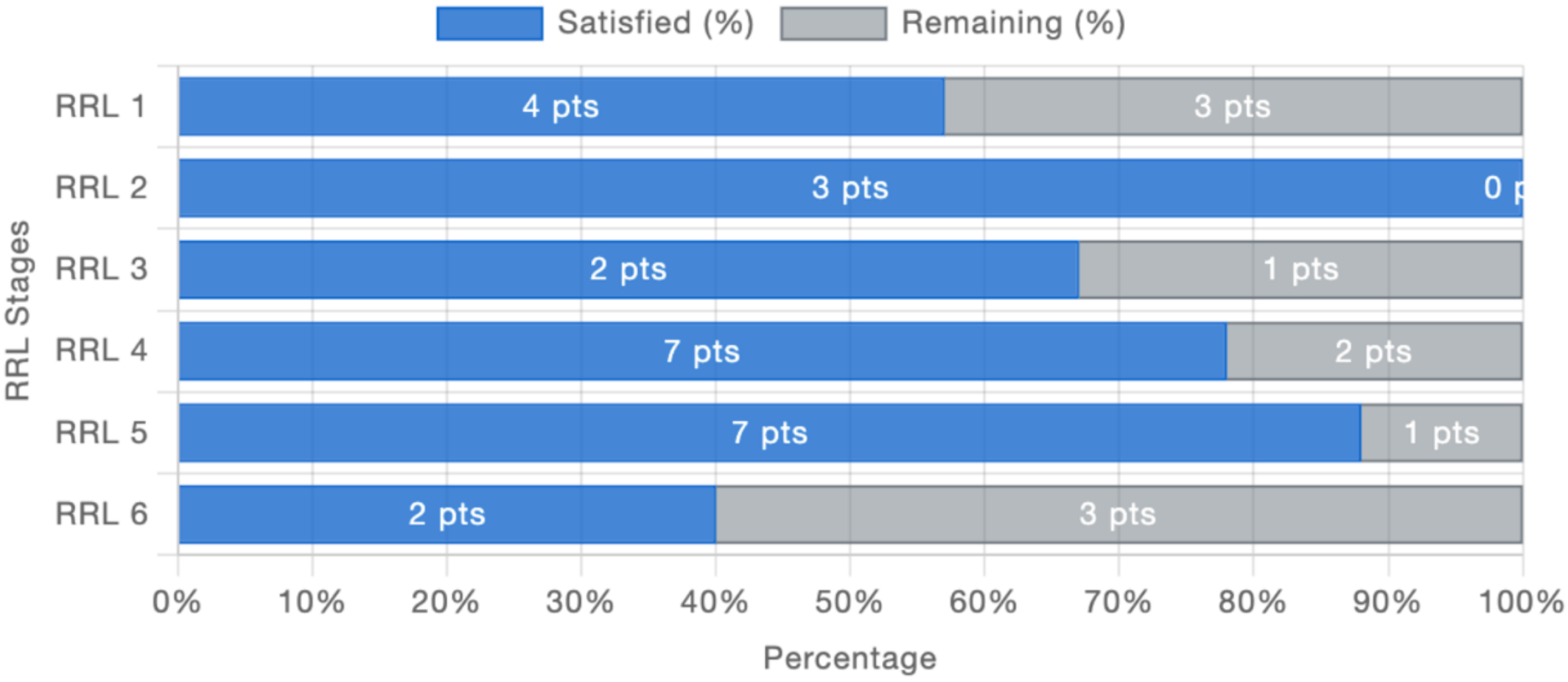

**Figure A3.** Points satisfied per stage. Stacked horizontal bars show, for each RRL stage, the percentage of points satisfied (blue) versus remaining (grey); in-bar labels indicate the number of satisfied points.

**Table A2** Detailed Radiomics Quality Score (RQS 2.0) evaluation for the foundation model-driven framework for rectal cancer classification. The study achieved an RQS 2.0 score of 25/35 (Radiomics Readiness Level 6), reflecting strong methodological transparency, validation, and interpretability consistent with the RQS 2.0 framework

| No. | Criteria | Selected Option | Points | Explanation |
|---|---|---|---|---|
| **RRL 1 - Foundational Exploration** | | | | |
| 1 | Unmet Clinical Need – Unmet clinical need (UCN) defined.<br>● UCN is agreed upon and defined by more than one center.<br>● UCN is defined using an established consensus method such as the Delphi method. | Implemented: More than one center (+1) | 1 | The study addresses the unmet need for reliable MRI-based detection of EVI and MFI in rectal cancer, agreed upon by three centres within the EU CHAIMELEON project. |
| 2 | Hardware Description – Detailed description of the imaging hardware used, including model, manufacturer, and technical specifications. | Implemented (+1) | 1 | Different MRI scanner manufacturers used across centres (Siemens, Philips, GE) are detailed in the manuscript table to document hardware variability. |
| 3 | Image Protocol Quality – Five levels of image protocol quality for TRIAC:<br>● Level 0: Protocol not formally approved.<br>● Level 1: Approved with a reference number in the institutional archive.<br>● Level 2: Approved with formal quality assurance (recommended minimum for prospective trials).<br>● Level 3: Established internationally; published in guidelines and peer-reviewed papers.<br>● Level 4: Future proof | Not implemented | 0 | The study used retrospectively collected multi-center MRI data (CHAIMELEON project) acquired under heterogeneous clinical protocols without a unified TRIAC-compliant acquisition standard. Although no formal acquisition QA was performed, frequency domain harmonization was applied to mitigate scanner-related variability. |

| | (follows TRIAC Level 3, FAIR principles, retains raw data). | | | |
|---|---|---|---|---|
| 4 | Inclusion and Exclusion Criteria – Detailed criteria for patient selection in studies, including rationale. | Implemented (+1) | 1 | Detailed inclusion/exclusion criteria were defined and are documented in the Methods (Dataset) section. |
| 5 | Diversity and Distribution – Identify potential biases before the project (demographics, socioeconomic, geographic, medical profiles). | Implemented (+1) | 1 | Patient demographics (age, gender) were summarised in Table 1. Data were collected from three hospitals in different regions, providing limited geographic and institutional diversity to reduce single-center bias. |
| **RRL 2 - Data Preparation** | | | | |
| 7 | Preprocessing of Images – Apply steps to standardize images with clear reasoning. | Implemented (+1) | 1 | Preprocessing steps are fully described in the Methods (Preprocessing) section. |
| 8 | Harmonization – Use image-level (e.g. CycleGANs) or feature-level (e.g. ComBat) harmonization techniques. | Implemented (+1) | 1 | A self-supervised frequency domain harmonization strategy was applied to reduce scanner variability across centres. |
| 10 | Automatic Segmentation – Use an automated segmentation algorithm for ROI definition. | Implemented (+1) | 1 | Rectum patch were automatically localised using the TotalSegmentator tool. |
| **RRL 3 - Prototype Model Development** | | | | |
| 13 | HCR + DL Combination – Compare and explore the synergistic combination of handcrafted radiomics and deep learning models. | Not implemented | 0 | Only deep learning–based features from the UMedPT framework were used; handcrafted radiomics features were not combined or compared. |
| 14 | Multivariable Analysis – Incorporate non-radiomics features (clinical, genomic, proteomic) to yield a holistic model. | Implemented (+2) | 2 | Age was integrated with the UMedPT_LR multiplanar model (axial+sagittal) as a clinical covariate. No significant improvement was observed (EVI: AUC 0.73 vs 0.82; MFI: AUC 0.71 vs 0.71). |
| **RRL 4 - Internal Validation** | | | | |
| 15 | Single Center Validation – Validation performed on data from the same institute without retraining or adapting the cut-off value. | Implemented (+1) | 1 | Data from three centers were combined and divided into training, validation, and test sets. |
| 16 | Cut-off Analyses – Identify optimal thresholds (e.g., using Youden's Index) for classification or survival analysis. | Implemented (+1) | 1 | A fixed probability threshold of 0.5 was used in line with CHAIMELEON challenge requirements. |
| 17 | Discrimination Statistics – Report discrimination metrics (e.g., ROC curve, sensitivity, specificity) with significance (p-values, CIs).<br>● Statistic reported<br>● With Resampling method | Resampling method applied (+2) | 2 | Discrimination metrics (ROC, AUC, sensitivity, specificity, 95% CI) were reported, and statistical significance assessed via resampling (e.g. bootstrapping). |
| 18 | Calibration Statistics – Report calibration metrics (e.g., calibration-in-the-large, slope, plots). | Implemented (+1) | 1 | Calibration plots are in the Supplementary materials. |
| 19 | Failure Mode Analysis – Document model limitations with examples of edge cases. | Implemented (+1) | 1 | Model limitations were discussed in the Discussion section, including reduced EVI performance with harmonisation and lower robustness of sagittal plane classification. |
| 20 | Open Science and Data – Make code and data publicly available.<br>● Open scans (+1)<br>● Open segmentations | One aspect (+1) | 1 | The code will be publicly available via GitHub repository: https://github.com/yumengzhang97/foundation-model-rectal-mri |
| | (+1)<br>● Open code (+1) | | | |
| | **RRL 5 - Capability Testing** | | | |
| 21 | Multi-center Validation – Validation with data from multiple institutes ensuring no overlap:<br>● One external institute<br>● Two or more external institutes<br>● Third-party platform with completely unseen data | Two or more institutes (+2) | 2 | Validation of the trained model was carried out on more than one centers. |
| 22 | Comparison with 'Current Clinical Standard' – Assess model agreement or superiority versus the current gold standard (e.g., TNM staging). | Implemented (+2) | 2 | Model performance was compared with the current clinical gold standard(histopathological assessment). |
| 23 | Comparison to Previous Work – Compare performance with published HCR signatures or DL algorithms. | Implemented (+1) | 1 | The proposed model was compared with previously the best performance of CHAIMELEON challenge. |
| 24 | Potential Clinical Utility – Report on the current and potential clinical application (e.g., decision curve analysis). | Implemented (+2) | 2 | The study discussed the model's potential clinical application in the discussion part. |
| | **RRL 6 - Trustworthiness Assessment** | | | |
| 25 | Explainability – Apply explainability tools (e.g., SHAP for HCR, GradCAM for DL) to clarify model predictions. | Implemented (+1) | 1 | Explainability was addressed using Grad-CAM visualizations. |
| 26 | Explainability Evaluation – Conduct qualitative and quantitative evaluations of interpretability methods (e.g., checking consistency to adversarial perturbations). | Implemented (+1) | 1 | Qualitative explainability analysis was performed using Grad-CAM visualizations to identify image regions contributing most to model predictions. The highlighted areas were visually compared with anatomically relevant regions for EVI and MFI to assess interpretability consistency. |
| 27 | Biological Correlates – Detect and discuss biological correlates to deepen understanding of radiomics and underlying biology. | Not implemented | 0 | No molecular or histopathological correlation analyses were performed. |
| 28 | Fairness Evaluation and Mitigation – Evaluate model performance for biases and apply bias correction if needed.<br>● Fairness evaluated<br>● Bias correction applied | Not implemented | 0 | Fairness evaluation was not performed, as subgroup analyses (e.g., by center, gender, or age) were beyond the scope of this study. Future work will include bias assessment across demographic and institutional subgroups. |
| | **Total = 25/35 (71%)** | | | |

## A5. Calibration curves

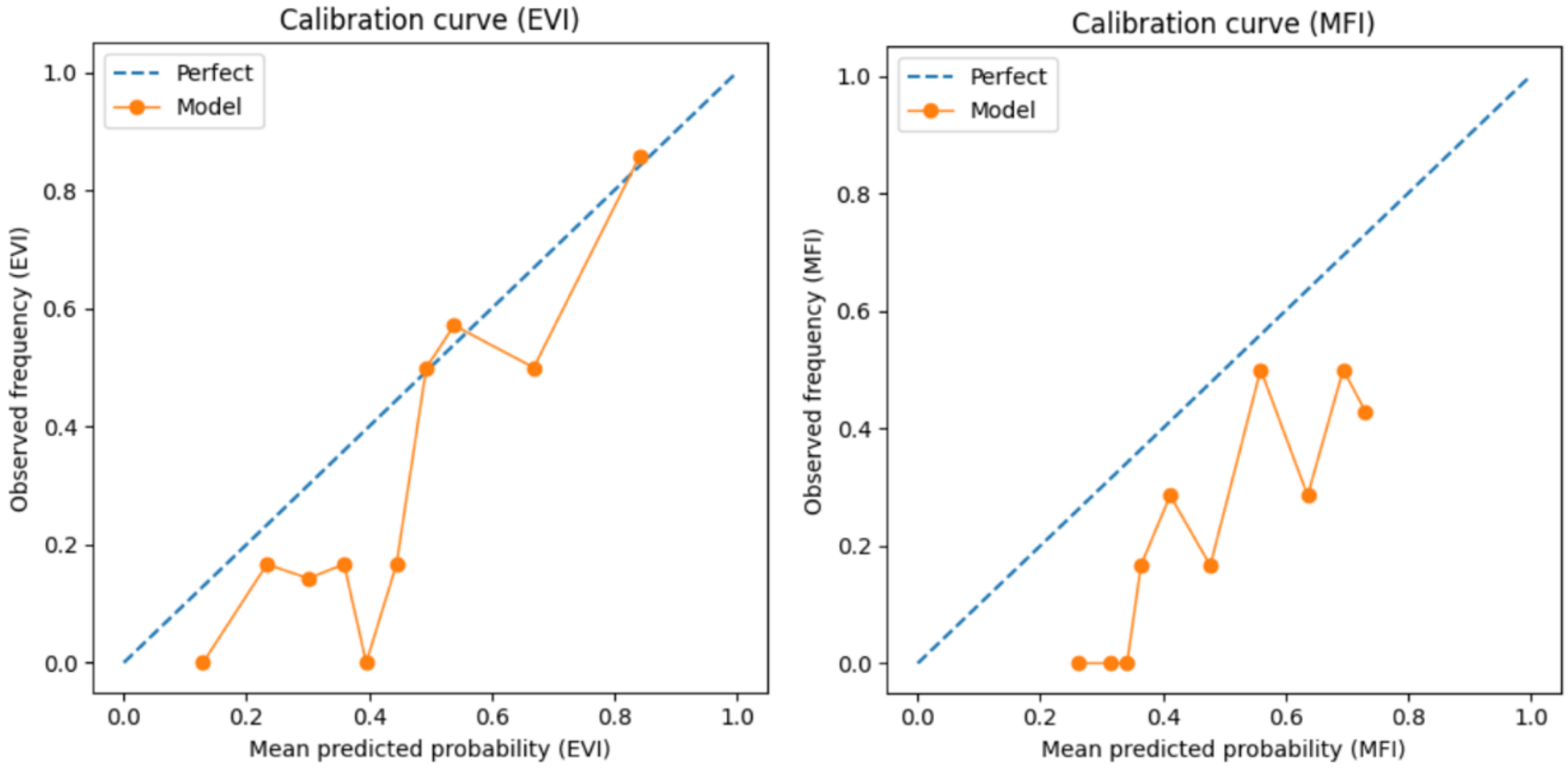

**Figure A4.** Calibration curves of the best EVI and MFI models on the test set (10 quantile bins). **EVI**—AUC = 0.82, Brier = 0.176, slope = 1.04, intercept = −0.69, ECE = 0.145; **MFI**—AUC = 0.77, Brier = 0.211, slope = 1.22, intercept = −1.28, ECE = 0.248.

## A6. Supplementary References

1. Huang H, Lin L, Tong R, et al (2020) UNet 3+: A full-scale connected UNet for medical image segmentation. arXiv [eess.IV]
2. Simonyan K, Zisserman A (2014) Very deep convolutional networks for large-scale image recognition. arXiv [cs.CV]
3. He K, Zhang X, Ren S, Sun J (2016) Deep residual learning for image recognition. In: 2016 IEEE Conference on Computer Vision and Pattern Recognition (CVPR). IEEE
4. Hu J, Shen L, Albanie S, et al (2017) Squeeze-and-Excitation Networks. arXiv [cs.CV]